\documentclass[proceedings, preprint]{rmaa}

\usepackage{paralist}

\usepackage{psfrag,color}

\SetYear{2008} \SetConfTitle{MFU II: from Laboratory and Stars to
 Primordial Structures}

\title{Supernova Explosions and the Triggering
of Galactic Fountains and Outflows}

\author{
  E. M. de Gouveia Dal Pino,\altaffilmark{1}
  C. Melioli,\altaffilmark{1,2}
  A. D'Ercole ,\altaffilmark{2}
  F. Brighenti,\altaffilmark{3}
  and A. Raga\altaffilmark{4}}

\altaffiltext{1}{Instituto Astronomico e Geofisico, Universidade de
Sao Paulo, IAG$-$USP, R. do Matao, 1226, Cd. Universitaria, Sao
Paulo, SP, Brazil (dalpino@astro.iag.usp.br).}

\altaffiltext{2}{INAF-Osservatorio Astronomico di Bologna, via
Ranzani 1, 40126 Bologna, Italy.}

\altaffiltext{3}{Dipartimento di Astronomia, Universita di Bologna,
via Ranzani 1, 40126 Bologna, Italy.}

\altaffiltext{4}{Instituto de Ciencias Nucleares,
  UNAM, Ciudad Universitaria, Apartado Postal 70-543, CP 04510, DF, M\'exico.}

\shortauthor{de Gouveia Dal Pino et al.}
\shorttitle{Galactic Fountains}

\listofauthors{E. M. de Gouveia Dal Pino, C. Melioli, A. D'Ercole,
F. Brighenti \& A. Raga}
\indexauthor{de Gouveia Dal Pino, E. M.}
\indexauthor{Melioli, C.}
\indexauthor{D'Ercole, A.}
\indexauthor{Brighenti, F.}
\indexauthor{Raga, A.}

\abstract{We review here the effects of supernovae (SNe) explosions
on the environment of star-forming galaxies. Randomly distributed,
clustered SNe explosions cause the formation of hot superbubbles
that drive either galactic fountains or supersonic winds out of the
galactic disk, depending on the amount and concentration of energy
that is injected by them. In a galactic fountain, the ejected gas is
re-captured by the gravitational potential and falls back onto the
disk. From 3D non-equilibrium radiative cooling hydrodynamical
simulations of these fountains, we find that they may reach
altitudes smaller than 5 kpc in the halo and hence explain the
formation of the so-called intermediate-velocity-clouds (IVCs) often
observed above the disk of these galaxies. On the other hand, the
high-velocity-clouds (HVCs) that are observed at higher altitudes
(of up to 12 kpc) require another mechanism to explain their
production. We argue that they could be formed either by the capture
of gas from the intergalactic medium and/or by the action of
magnetic fields that are carried out to the halo with the gas in the
fountains. Due to angular momentum losses (of 10-15\%) to the
halo, we find that the fountain material falls back to smaller radii
and is not largely spread over the galactic disk, as previously
expected, but falls near the region where the fountain was produced
This result is consistent with the metal distribution derived
from recent chemical models of the galaxy. We also find that after
about 150 Myr, the gas circulation between the halo and the disk in
the fountains reaches a steady state regime, and this is relatively
insensitive to the galacto-centric distance where the fountains are
produced. The fall back material leads to the formation of new
generations of complex structures that help to feed the supersonic
turbulence in the disk. }

\resumen{Damos una rese\~na de los efectos de explosiones de
supernovas (SNe) sobre el medio ambiente de galaxias con formaci\'on
estelar. Explosiones de SNe en c\'umulos distribuidos al azar
producen super burbujas calientes que impulsan fuentes gal\'acticas
o vientos supers\'onicos que salen del disco gal\'actico, dependiendo
de la cantidad y concentraci\'on de la energ\'\i a que inyecten. En
una fuente gal\'actica, el gas eyectado es recapturado por el
potencial gravitacional, y vuelve a caer sobre el disco. Con simulaciones
3D de din\'amica de gases radiativos fuera de equilibrio de estas
fuentes gal\'acticas, encontramos que pueden llegar a alturas menores
a 5 kpc en el halo, y de esta manera explicar las nubes de velocidad
intermedia (IVCs) frecuentemente observadas sobre los discos de estas
galaxias. Por otro lado, las nubes de alta velocidad (HVCs) observadas
a mayores alturas (hasta de 1 kpc) requieren de otro mecanismo para
explicar su producci\'on. Argumentamos que podr\'\i an ser formadas
por captura de gas del medio intergal\'actico y/o por la acci\'on
de campos magn\'eticos llevados hacia el halo con el gas de las
fuentes gal\'acticas. Debido a p\'erdidas de momento angular
(de 10-15\%) al halo, encontramos que el material de las fuentes
cae sobre radios menores, y no es esparcido
sobre el disco, como era esperado, cayendo en vez cerca de la regi\'on
donde est\'a la fuente. Este resultado es consistente con la
distribuci\'on de metales de recientes modelos qu\'\i micos de la galaxia.
Tambien encontramos que despues de aproximadamente 150 Myr la
circulaci\'on del gas entre el halo y el disco en las fuentes
llega a un estado estacionario, el cual es poco sensible a la
distancia de las fuentes al centro gal\'actico. El material que
vuelve a caer sobre el disco lleva a la formaci\'on de nuevas
generaciones de estructuras complejas que alimentan la turbulencia
supers\'onica del disco.}

\addkeyword{Galaxy: galactic fountains} \addkeyword{Galaxy: winds}
\addkeyword{ISM: supernovae explosions}  \addkeyword{ISM:
superbubbles}

\begin{document}
\maketitle

\section{Introduction}
\label{sec:intro}

Edge-on star forming disk galaxies often exhibit hot halos of
ionized gas that may extend for several kpc above the regular HI
galactic disk. They are fed by ascending gas from the disk in
structures that resemble chimneys and fountains. Observations
indicate that the chimneys are generated by supernovae (SNe)
explosions which blow superbubbles that expand and carve holes in
the disk, injecting high speed, metal enriched gas that forces its
way out through relatively narrow channels with widths of 100 - 150
pc. They establish a connection between the thin disk and the halo,
feeding it with the hot disk gas that expands under buoyancy forces
up to a maximum height into the halo and then returns to the disk,
pulled by the disk gravity. The whole cycle is like a fountain - hence
the name galactic fountain. Evidence for large chimneys is clear in
external galaxies in the form of holes and flows in the distribution
of HI. In the Milky Way (MW), the evidence has mainly been in the
form of fragments and vertical structures in the large scale maps of
the interstellar medium. A multi-wavelength survey of the halos of
several star forming galaxies (e.g., \citet{det05}; Dettmar, these
Proceedings) have revealed a correlation of these halos with the
rates of star formation and the energy input rates by SNe, suggesting
that gaseous halos are associated to star formation processes in the
disk.

Other observed features that seem to be correlated to gas
circulation in chimneys and galactic fountains are the so-called
intermediate and high-velocity-clouds (IVCs and HVCs, respectively).
These are mainly neutral hydrogen (HI) clouds that can be as large
as 100 pc, with masses of up to 10$^4$M$_\odot$ that are observed in
the halo of the MW and other star forming galaxies at altitudes
typically between 300 pc and 2.5 kpc, which are falling on the disk
with velocities between -20 km/s and -90 km/s. The HVCs can be
observed at even higher altitudes (up to 12 kpc) and with velocities
of up to -140 km/s. Figure 1 provides a mosaic of the matter
distribution in the halo of the MW (see also \citet{gou08}) which
shows that the galactic disk is surrounded by a very cloudy
environment. It is generally believed that at least the IVCs have
been formed from the condensation of the gas that arises in the
chimneys triggered by SN explosions, and numerical simulations seem
to confirm this hypothesis (e.g., \citet{avi00}, \citet{avi05},
\citet{mel08a}, \citet{mel08b}, see below). However, the origin of
the HVCs is still controversial. The difficulty at producing
fountains (in hydrodynamical simulations) reaching altitudes higher
than 5 kpc \citep{mel08b} and the very small metallicity contents
observed in these HVCs suggest that they may have originated from
gas raining into the galaxy, accreted from the intergalactic medium
(IGM) or from satellite galaxies (e.g., \citet{fra06}).

\begin{figure}[!t]
  \includegraphics[width=\columnwidth]{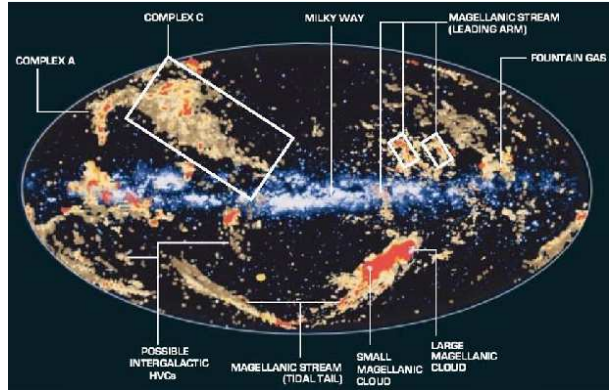}
  \caption{Map of the galactic gas and its environment that
   combines radio observations of neutral hydrogen (HI) of the environment
   with a visible light image of the MW (the galactic disk is in the middle).
   The high and intermediate-velocity clouds of hydrogen, such as complexes
   A and C, are located above and below the disk.
   A galactic fountain is also identified in the map
   (extracted from http://www.astro.uni-bonn.de/~webiaef/outreach/posters/milkyway/).}
  \label{fig:simple}
\end{figure}

Other examples of violent gas outflows coming out of  the disk of
star forming galaxies are the supersonic winds, which are powerful
enough to escape from the gravitational potential of the galactic
disk to the IGM. A large-scale bipolar wind seems to be emerging
from the center of the MW \citep{bla03} and spectacular winds
extending for several kpc above the disk have been observed in
galaxies with bursts of star formation. These starburst (SB)
galaxies are actually merging or interacting galaxies that can have
star formation rates up to 20 times larger than those of regular
galaxies, and this could explain the large amount of energy
injection by SNe and the resulting production of powerful winds.
Galactic winds are indeed ubiquitous in SBs (\citet{vei05} and
references therein). Recent, high resolution observations of the
best studied prototype of this class, the Starburst galaxy M82, show
evidence that its wind, as in galactic fountains, is being fed by
SNe explosions nestled in several stellar associations around the
nuclear region of the galaxy \citep{kon08}. In this brief review, we
will focus on the formation of galactic fountains and thus will not
address any further these powerful galactic winds, but we refer to,
e.g., \citet{mel04} and \citet{cop08} (and references therein) for a
more complete discussion of the efficiency of the SNe in powering
the SBs and their outflow production.

Extensive work on the formation of galactic chimneys and fountains
has been carried out over the last decades (see e.g., de Avillez
2000; de Avilles \& Breitschwerdt 2005; Melioli et al. 2008a, and
references therein for reviews). \citet{sha76} first
proposed the idea that galactic chimneys induced by SNe explosions
would cause gas circulation between the disk and the halo. This
scenario was afterwards explored analytically in detail by \citet{bre80}
and \citet{kah81}. More recently, the advent of powerful
computers have made it possible to simulate fountains and winds
rather accurately. The first 3D hydrodynamical simulations following
the whole cycle of the gas between the disk and the halo in
fountains in the MW were performed by de \citet{avi00} and de
\citet{avi01}. However, in order to obtain a high spatial
resolution, they considered only a small region of the Galaxy with a
dimension of 1 kpc$^2$ (in the disk) $\times$ 10 kpc (for the height
in the halo). \citet{kor99} have included the effects of the
differential rotation and the magnetic field of the galactic disk,
but considered a computational domain too small (500 pc$^2$ x 1 kpc)
to allow for the development of an entire cycle of the chimney gas
between the disk and the halo. Other works that were also concerned
with the collective effects of supernovae on the structure of the
interstellar medium (ISM) have considered even smaller volumes in
their hydrodynamical simulations (see, e.g., \citet{mac89}).
More recently, further MHD simulations were carried
out by \citet{avi05}, without including
differential rotation, where again, in order to reach very high
resolution (of 0.6 pc) they considered only a small volume of the
Galaxy (1 kpc$^2 \times$  10 kpc), as they were primarily concerned
at examining the role of the disk-halo gas circulation in
establishing the volume filling factors of the different phases of
the ISM in the Galactic disk. In particular, these MHD studies have
revealed that the gas transport into the halo is not prevented by
the parallel magnetic field of the Galaxy (as suggested in former
works; see e.g., \citet{tom98}), but only delayed by few tens of Myr
when compared to pure HD simulations.

In the following sections, we summarize the results of a recent study
of the large scale development of galactic fountains driven by SNe
explosions, which was carried out in order to understand their
dynamical evolution, the observed kinematics of the extraplanar gas,
the formation of IVCs and HVCs, and the influence of the fountains in
the redistribution of the freshly delivered metals over the galactic
disk. To this aim, we have performed fully 3D, non-equilibrium
radiative cooling hydrodynamical simulations of the gas in the Milky
Way, in which the whole Galaxy structure, the galactic differential
rotation, the total (thermal $+$ magnetic) pressure, and the
supernovae explosions generated in single and multiple stellar
associations of OB stars have been considered (see Melioli et al.
2008a, 2008b for more details). We will also discuss qualitatively
the  dynamical influence that the magnetic fields are expected to
have on these outflows.

\section{Numerical Simulations}

The initial numerical setup is described in detail in Melioli et al.
(2008a, 2008b). The ISM is initially set in rotational equilibrium
in the galactic gravitational potential, given by the addition
of the dark-matter halo, the bulge, and the disk contributions.
Assuming hydrostatic equilibrium in the $z$- (vertical) direction
between gravity and the full pressure (given by the thermal gas $+$
magnetic $+$ cosmic ray pressure contributions), we have computed the
rotation velocity as a function of the galacto-centric distance $R$.
This allowed us to build the full galactic rotation curve, as
observed in the MW for different $z$ heights. The ISM in our model is
made up of the three gas components, namely the molecular (H$_2$),
the neutral (H I) and the (H II) hydrogen components, and their stratified
density distribution follows the empiric curves obtained for the MW
by \citet{wol03}. In most models, a hot ($T_h$ = 7 $\times$
10$^6$ K) isothermal gas halo was added, in equilibrium with
the galactic potential well. In some models the halo was allowed to
rotate with a velocity that was a fraction of the disk velocity.

In order to drive the formation of chimneys and fountains, we have
considered two sets of models. In one set, we continuously exploded
100 SNe over 30 Myr in a single star cluster. In a second set, we
randomly exploded up to 2000 SNe in multiple star clusters (spread
over an area of either 1 kpc$^2$ or 8 kpc$^2$ on the disk). In both
cases, the clusters were generally localized at a radial distance
$R=$8.5 kpc, which corresponds to the distance between the Sun and the galactic
center. We have assumed the SNe rate distribution over time that has
been inferred from observations of the MW \citep{hig05}.

The simulations  employ a modified version of the adaptive mesh
refinement  YGUAZU code that integrates the 3D invisced gasdynamic
equations with the flux vector splitting algorithm of van Leer (\citet{rag00};
\citet{rag02}; see also, e.g., \citet{gon04}; \citet{mel05}).
The non-equilibrium radiative
cooling of the gas is computed together with a set of continuity
equations for atomic/ionic or chemical species. The 3D binary,
hierarchical computational grid has a uniform base grid, and
a number of higher resolution grids at chosen spatial positions.
We have enforced the
maximum grid resolution only in the volume encompassed by the
fountain. In order to follow the circulation and the thermal history
(i.e., the degree of radiative cooling) of the metals expelled by the
SNe, we have added three different tracers passively advected by the
code describing the disk gas, the halo gas, and the SNe ejecta (see
Melioli et al. 2008a for details).

Figures 2 and 3 exhibit the face-on and edge-on views of the
evolution of a galactic fountain arising from SNe explosions within
a single OB association located at the galacto-centric distance $R
=$ 8.5 kpc. At this radius the transition between the disk and the
hot halo occurs at $z =$ 800 pc. The critical luminosity for the SNe
to break through the disk is $L_b \simeq 1.5 \times 10^{37}$ erg/s
(e.g., \citet{koo92}), which is well below the mechanical
luminosity $L_W = 10^{38}$ erg/s provided by the 100 SNe powering
the fountain. This model was run with a maximum spatial resolution
of 12.5 pc.

During its activity, the fountain digs a hole in the disk and throws
SNe ejecta and ISM vertically up to $z \sim$ 2 kpc above the
galactic plane. Once the  explosions cease, the hole collapses in $2
\times 10^7$ yr and the ejecta trapped at its edges (nearly half of
the total) mixes with the local ISM. Owing to the differential
rotation of the Galactic disk, the ejecta is not confined to one
spot, but is stretched, giving rise to the bean-like structure seen
in Figure 2. The low density tail with a banana-shape is due to the
ejecta pushed at high altitudes which then slowly falls back, remaining
above the disk. The two concentric circles in Figure 2 with radii $R
=$ 8 kpc and $R =$ 9 kpc, have been drawn to guide the eye. The SNe
explode between these two circles, and their distance corresponds to
the diameter of the hole produced by the fountain. Comparing the
shape of the tail traced by the ejecta with the circles, we note
that the gas of the fountain tends to move inward during its
trajectory. Actually, as the gas moves upward and interacts with the
halo, it transfers to it part of its angular momentum; the
centrifugal force decreases in pace with the circular velocity and
gravity prevails, pushing the gas toward the Galactic center (see Figure 3).

\begin{figure}[!t]
  \includegraphics[width=\columnwidth]{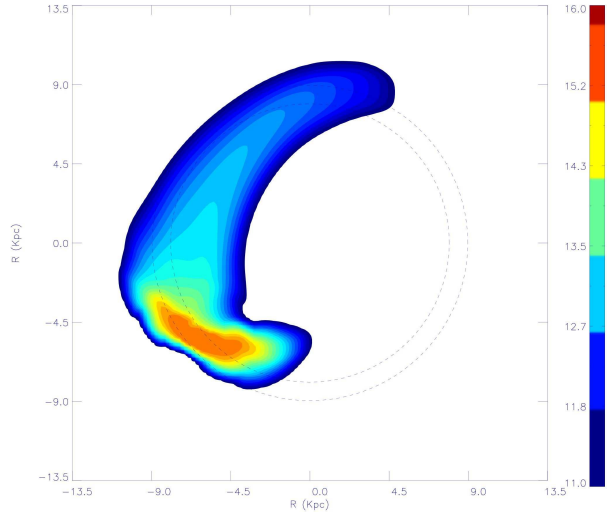}
  \caption{Face-on view of the $z$-column density distribution of the
  SN ejecta after $t =$ 160 Myr for a single fountain.
  The SNe explode at a radius $R =$ 8.5
  kpc, halfway between the radii $R =$ 8 kpc and $R =$ 9 kpc of the two circles
  drawn in the figure to guide the eye.
  Their distance represents approximately the maximum extension of the hole
  in the ISM carved by the fountain on the disk during its activity. At the
  depicted time, the
  hole has collapsed and disappeared.
  The column density scale is given in cm$^{-2}$ and the $x,y$ axes are
  labeled in kpc (from Melioli et al. 2008a).}
  \label{fig:2}
\end{figure}

\begin{figure}[!t]
  \includegraphics[width=\columnwidth]{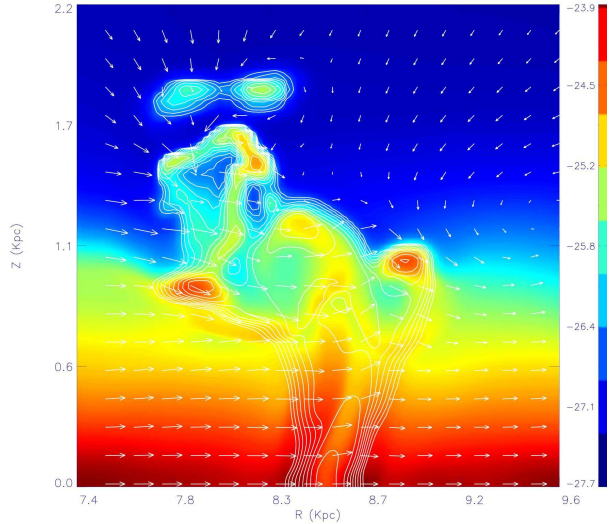}
  \caption{Edge-on view of the single fountain of figure 1 at $t=$ 50 Myr,
  when the
  ascending gas reaches its maximum height ($z = 2$ kpc).
  Isodensity curves of the ejecta
  are superimposed on the disk gas density distribution,
  highlighting the fountain pattern and the cloud formation
  (see the text; from Melioli et al. 2008a).)}
  \label{fig:3}
\end{figure}

Further insight on the evolution of the galactic fountain is
obtained from Figure 4, which illustrates several quantities starting
from $t= 30$ Myr, the time at which the SNe stop exploding. The upper
panel shows how quickly the ejecta looses its angular momentum
because of the interaction with the gaseous halo. After 80 Myr,
nearly 10\%  of the angular momentum of the fountain has been
transferred to the hot halo; later on, no further transfer occurs
because nearly 75\% of the ejecta is located below the disk-halo
transition and rotates together with the ISM. The middle panel of
Figure 4 shows the ejecta mass fraction located above different
heights. It is interesting to note that for $t > 80$ Myr, the
long-term evolution of these quantities becomes very slow. This is
due to the fact that most of the ejecta situated above the plane is
rather diluted and tends to float together with the extra-planar
ISM. Finally, in the lower panel of Figure 4, we quantify the
tendency of the ejecta to move radially by plotting the fraction of
the total mass of ejecta located at $R < 9$ kpc and $R > 8$ kpc
(i.e. the radii of the two circles drawn in Figure 2); the amount of
mass located within the region $8 \le R \le 9$ kpc is not taken into
account. In the beginning, the ejecta starts to follow the expected
tendency to move outward but, as the loss of angular momentum
proceeds, the fraction of gas moving inwards increases, and after 60
Myr overrides the outward directed mass transport.

\begin{figure}[!t]
    \includegraphics[height=10cm]{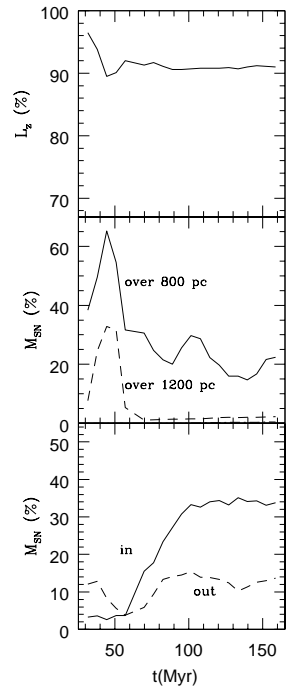}
  \caption{The upper panel shows the evolution of the angular momentum of the SN
  ejecta for the single galactic fountain of Figures 2 and 3.
  The evolution of the amount of the ejecta above different heights is illustrated
  in the middle panel. The lower panel displays the temporal behavior of the amount
  of ejecta located at $R < 8$ kpc (dashed line) and $R > 9$ kpc
  (solid line); these two radii correspond to the circles visible in Figure 2
  (from Melioli et al. 2008a).}
  \label{fig:4}
\end{figure}

The gas lifted up by the fountain has a mass of $2.5 \times 10^5$M$_\odot$,
almost all (92 \%) condensed in dense filaments cooled to $T
= 10^4$ K (see Figure 3). The clouds form via thermal instabilities
at $z \sim 2$ kpc, the maximum height that the ascending gas reaches
before starting to move back toward the disk at 50 Myr. All the clouds
have negative $z$-velocities in the 50-100 km/s range. The
chemical composition of these clouds is practically unaffected by
the SN ejecta. The fountain is powered by 100 SNe, half of them
exploding in the half space mapped by the grid; as each supernova
delivers on average 3~M$\odot$ of metals, a total of 150~M$_\odot$
of heavy elements is ejected by the fountain in Figures
2 and 3. Only 20\% of the metals of the ejecta is locked in the
clouds, 50\% of this material remains trapped within the disk and
30\% remains floating over the disk as hot, diffuse gas. As a
result, the metallicity increment in the clouds due to the freshly
delivered metals corresponds to 0.01 in solar units and is
negligible compared to the solar abundance of the ISM. As we will
show below, this result also holds in models with multiple SNe
associations. In conclusion, almost all the gas lifted up by the
fountain condenses into clouds without being chemically affected.
After 150 Myr,  45\% of the fresh metals stays on the disk (below $z
=$ 800 pc) within a radial distance of $R =$ 0.5 kpc from the OB
association. A  fraction of 35\% is found on the disk within the
$9.5 < R < 7$ kpc range. The remaining 20\% of metals is still over
the disk, half of which is at $R >$ 8.5 kpc and half at $R < $ 8.5 kpc.

Figure 5 shows the results for a multiple fountain model that was
produced by randomly clustered explosions of SNe originating in
stellar associations spread over a disk area of 8 kpc$^2$ for a
period $P=$ 200 Myr with a mean rate adequately scaled from the
rate of $1.4 \times 10^{-2}$  yr$^{-1}$ for the whole Galaxy
\citep{cap97}. During the time $P$ of the simulation, $N_{tot} = 1.59
\times 10^4$ SNe exploded in the active area of the Galactic disk.
The employed size-frequency distribution of the clustered SNe
progenitors follows the distribution inferred from observations by
\citet{hig05} for the MW.

\begin{figure}[!t]
\includegraphics[width=\columnwidth]{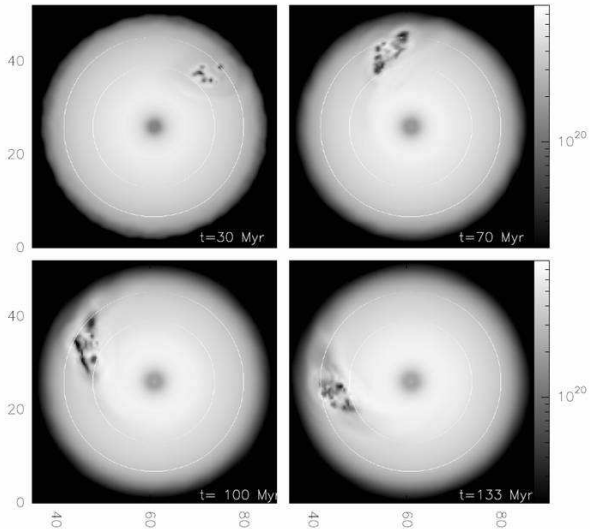}
  \caption{Face-on view of the evolution of a multiple fountain
  triggered by the explosion of SNe from randomly distributed stellar clusters
  over an area of
  8 kpc$^2$ of the galactic disk at a distance $R=$ 8.5 kpc from the galactic center.
  The figure shows 4 snapshots of the evolution of the column density (in cm$^{-2}$)
  of the ascending gas (ISM$+$SNe ejecta) in a multiple fountain (from Melioli et al. 2008b).}
  \label{fig:5}
\end{figure}

We may notice the holes dug by the SNe into the disk in the
beginning, at $t=$ 30 Myr. As time goes by, as in the case of the
single fountain (of Figures 2 and 3), the ascending material in the
multiple fountains is stretched by the rotation, forming cometary
tail-like structures. Again, as the ascending material returns to the
disk it tends to fall towards the inner disk region due to angular
momentum losses (of about 15\%) to the halo. In this case, the
maximum height attained by the fountain material, with the formation
of dense cold clouds from the condensation of the hot gas, is two times
larger than in the case of the single fountain, and the clouds rain
back on the disk with velocities between $-50$ km/s and $-100$ km/s.
Hence, these results indicate that the galactic fountains are able
to produce only IVCs. We have also found that these results are
relatively insensitive to the distance of the fountains to the
galactic center (they were also simulated at a distance of 4.5 kpc
from the galactic center). Our simulations of multiple fountains
have also revealed that the amount of gas circulating between the
halo and the disk reaches a dynamical equilibrium around 150 Myr, so
that after this time the amount of gas falling back on the disk is
approximately equal to the gas ascending in the fountain (Figure
6). As before, the spreading of the SNe ejecta falling back on the disk is
not very large. Most of the gas lifted by the fountain falls back
within a distance $\Delta R = \pm$ 0.5 kpc from the place where the
fountain was originated. This is in agreement with recent chemical
models of the metal distribution in the MW disk \citep{ces07}.

\begin{figure}[!t]
 \includegraphics[height=11cm]{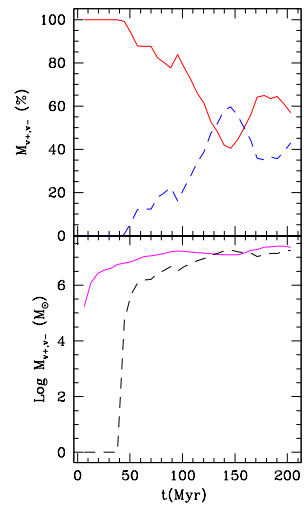}
  \caption{Disk-halo gas circulation in the multiple fountain model of Figure
  5. The amount of gas going up (continuous line) and down (dashed line)
reaches a  dynamical equilibrium around 150 Myr
 (from Melioli et al. 2008b).}
  \label{fig:6}
\end{figure}

\section{Conclusions and  discussion}

Large scale 3D hydrodynamical, non-equilibrium radiative cooling
simulations of random explosions of SNe in off-center stellar
clusters in a rotating galactic disk-bulge system evidence the
formation of giant superbubbles that break through the galactic disk
into the halo, forming chimneys and fountains:
\begin{asparaitem}
\item The gas lifted by multiple fountains condenses into clouds
that reach altitudes smaller than 5 kpc and then fall back into the
disk with velocities between $\sim -50$ km/s and $-100$ km/s.
These clouds can thus explain the formation of the
IVCs, but not of the HVCs that are observed at heights as large as
12 kpc in the halo. After about 150 Myr, the gas circulation between
the halo and the disk in the fountains is found to reach a steady
state regime, which is relatively insensitive to the
galacto-centric distance where the fountains are produced.

\item After a maximum lift, the clouds rain back on the disk, but
towards smaller radii due to angular momentum losses (10\%-15\%) to
the halo. This amount of angular momentum may be large enough to
provoke an eventual co-rotation of the halo with the disk. However,
the observed halos of star forming galaxies often show a rotational
velocity gradient with respect to the disk of $\Delta v_{rot} \simeq
-$15 km/s/kpc (for $1.3 < z < 5.2$ kpc; \citet{det05}; see also
Detmar in these Proceedings). This could be an indication that some
other mechanism (possibly of external origin) might be operating to
inhibit the halo rotation (see more details in \citet{mel08b}).

\item The galaxy rotation inhibits both a straight vertical expansion of
the fountains and the spreading of the metals transported by the
fountains from the SNe. Contrary to ballistic models, most of the
gas that is lifted up by the fountains falls back on the disk within
a distance $\Delta R = \pm 0.5$ kpc from the place where the
fountain originated. As a consequence, nearly 60\% of the metals
delivered by the SNe remains within the area where the fountain was
formed. This small radial displacement of metals in the disk is in
agreement with recent chemical models of the Milky Way (Cescutti et
al. 2007).

\item We have also found from the simulations that the metal pollution of
the clouds is $\sim$ 0.01 in solar units, therefore, negligible
compared to the typical ISM abundance of metals. This implies that
the clouds that are formed by the fountains in the halo are only very
weakly chemically affected by the SNe.

\item Once the material of the clouds rains back onto the disk, we may expect
that it will provide the formation of new molecular clouds and
supersonic filamentary structures that will feed the ISM turbulence,
thus closing the gas cycle between the disk and the halo (e.g., de
Avillez \& Breitschwerdt 2005).

\item We have also performed hybrid simulations including both the ejected
material from fountains triggered by SNe arising within the galactic
disk and the infall of gas  accreted from the IGM at a $dM/dt =1$ to
2$M_\odot$yr$^{-1}$ rate (see Figure 7). In this case, the
formation of low-metallicity, high-velocity halo structures falling
on the disk from the highest latitudes (like the HVCs) is a
straightforward consequence.
\end{asparaitem}

\begin{figure}[!t]
  \includegraphics[width=\columnwidth]{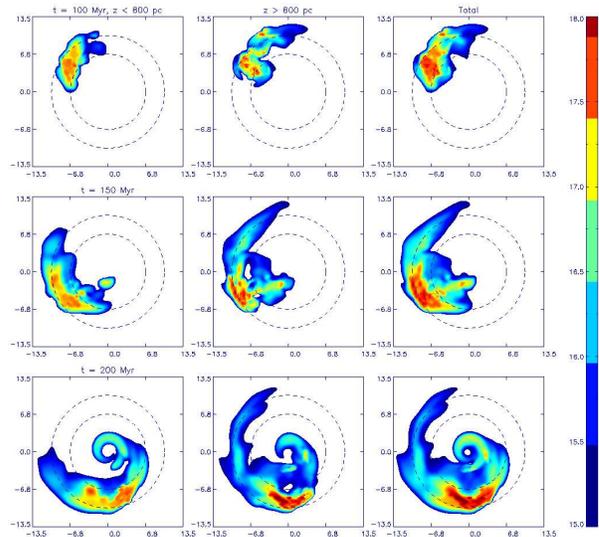}
  \caption{Multiple fountain $+$ IGM combined matter infall.
  Face-on view of the evolution of the multiple fountain
  triggered by the explosion of SNe from randomly distributed stellar
  clusters over an area of
  8 kpc$^2$ of the galactic disk at a distance $R=$ 8.5 kpc from the galactic
  center,
   as in Figure 5, but including infalling gas accreted from the IGM at
   a $dM/dt = 1\to 2$M$_\odot$yr$^{-1}$ rate.
The figure shows 3 snapshots of the evolution up to 200 Myr of the
column
  density distribution (in cm$^{-2}$)
  of the  (ascending $+$ infalling) gas (top panel t= 100 Mry; middle panel: 150 Myr; and bottom panel: t = 200 Myr).
  The left and central panels correspond to the column densities
calculated in the intervals $z < 800$ pc and $z > 800 $ pc,
respectively, where $z = 800$ pc corresponds to the location of the
disk/halo transition. The right column shows the total column
densities. In this case we find the formation of
  high-velocity structures at heights larger than 5 kpc due to the
  infall (from Melioli et al. 2008b).}
  \label{fig:7}
\end{figure}

A final remark is in order. In the study reviewed here, we have
included all the essential ingredients of a star-forming disk
galaxy, except for the magnetic fields and the thermal conduction
effects. The magnetic fields may be significant at helping to lift
the fountains and forming, e.g., at least part of the population of
HVCs. In fact, large scale magnetic fields with coherent scales of
$\sim$ 1kpc have been observed in the halos of several of these
galaxies (e.g., Dettmar 2005; see also Detmar, these Proceedings).
The MHD simulations of galactic fountains by de Avillez \&
Breidtschwerdt (2005) have produced halo magnetic fields, but with
smaller scales. MHD effects have also been invoked by \citet{gou99}
to accelerate SNe-triggered winds in SB galaxies by a
magnetocentrifugal process and more recently, \citet{otm07} (see
also Otmianowska-Mazur et al. in these Proceedings) have
investigated the joint action of both the Parker-Rayleigh-Taylor
instability in a galactic dynamo and the production of cosmic rays
by SNe to explain the ascension of large scale magnetic fields to
the halo. These effects could also provide guidance and acceleration
for the fountains and help the formation of clouds at higher
latitudes than those allowed by purely hydrodynamical simulations.
Concerning the effects of thermal conduction, these can be highly
important at suppressing thermal instabilities and thus structure
formation, particularly at the smallest scales (not addressed in the
present study). On the other hand, conduction might be inhibited by
stochastic magnetic fields. These important ingredients will be
considered in future work.

\end{document}